\documentclass[useAMS,usenatbib]{mn2e}
\usepackage{graphicx} 
\usepackage{aas_macros}

\newcommand{\degree}{\ensuremath{^\circ}}

\title[On the origin of eccentricities among extrasolar planets]
{On the origin of eccentricities among extrasolar planets} 
\author[Daniel Malmberg and Melvyn B. Davies]
{Daniel Malmberg$^{1}$\thanks{E-mail:danielm@astro.lu.se 
(DM); mbd@astro.lu.se (MBD)} and Melvyn B. Davies$^{1}$  \\ 
$^{1}$Lund Observatory, Box 43, SE--221 00, Lund, 
Sweden }

\begin{document}
\date{Accepted for publication}
\pagerange{\pageref{firstpage}--\pageref{lastpage}} 
\pubyear{2008}
\maketitle
\label{firstpage}

\begin{abstract}
Most observed extrasolar planets have masses similar to, but orbits very different from, the gas giants 
of our solar system. Many are much closer to their parent stars than would have been expected and 
their orbits are often rather eccentric. We show that some of these planets might have formed in 
systems much like our 
solar system, i.e. in systems where the gas giants were originally on orbits with a semi-major axis of 
several au, but where the masses of the gas giants were all rather similar. If such a system is perturbed 
by another star, strong planet-planet interactions follow, causing the ejection of several planets while 
leaving those remaining on much tighter and more eccentric orbits. The eccentricity distribution of 
these perturbed systems is very similar to that of the observed extrasolar planets with semi-major axis 
between 1 and 6 au.
\end{abstract}


\section{Introduction} \label{sec:intro}
Since their first discovery \citep{1992Natur.355..145W,1995Natur.378..355M}, more than 300 
extrasolar planets have been found. In Fig. 1 we plot the semi-major axes and eccentricities of the 
planets detected using the radial velocity method as of October, 2008 \citep{2006ApJ...646..505B,
2008exoplanet.eu, 2008A&A...480L..33T}. As can be seen from the figure, the spread in eccentricity 
and separation is very large. 
A majority of the detected extrasolar planets have masses similar too, or larger than, those of the gas giants in our
solar system. Most planet formation models predict that such massive planets can only 
form outside the so-called snow line (situated at 3 au around a solar-mass star \citep{1998Icar..131..171K}) 
but as can be seen from Fig. 1, most of the detected exoplanets have orbits tighter
than this value. Thus, the orbits of most of the observed exoplanets must have shrunk considerably since 
their formation. The most efficient mechanism behind this is widely believed to be disk migration 
\citep{1996Natur.380..606L}. However, disk migration almost exclusively produce planets on 
{\sl circular} orbits, which does not agree with observations. Thus, in order to explain the observed 
eccentricities an additional mechanism must be at work.

The most popular such mechanism is scatterings due to strong planet-planet interactions which can explain 
the observed eccentricities. Several different models which reproduce the 
observed eccentricities very well have been suggested and most likely the observed sample 
is created from a combination of these and perhaps others.
It may, for example, be that planets come very close to each other while 
undergoing migration in the disk \citep{2005Icar..178..517M} or that the planetary orbits
are kept stable by the disk and as it evaporates the system becomes unstable \citep{2008ApJ...675.1538T}.
Another possibility is that many planetary systems are initially too tightly packed, leading
to that they become unstable on a time-scale of millions to several hundred million years and
undergo strong planet-planet interactions \citep{2002Icar..156..570M,2004ApJ...611..494B,
2005Natur.434..873F,2007astro.ph..3166C, 2007astro.ph..3160J}. 

It is however also possible that strong planet-planet interactions are triggered in long-term stable 
planetary systems (like, for example, our own solar system) by external perturbations. Two examples
of such are nearby passing stars in young stellar clusters
\citep{2004AJ....128..869Z,2006ApJ...641..504A,2007MNRAS.378.1207M} and the effects of a  
stellar companion in  a binary \citep[see for example][]{1997Natur.386..254H,1997ApJ...477L.103M, 
1999AJ....117..621H}.
In the latter case there are two distinctly different scenarios. If the planetary system formed in a primordial
binary system, the orbits of the planets and the companion star are expected to be essentially co-planar. 
The evolution of such a system has been studied extensively in, for example, \citet{
2005ApJ...618..502M}. If the planetary system instead formed around an originally single star, 
which was later exchanged into a binary in an encounter in a young stellar cluster, the orientation
of the orbits of the planets is completely random with respect to the orbit of the companion star. The
evolution of the system is then very different from the co-planar case \citep{2005ApJ...627.1001T,
2007MNRAS.377L...1M,2007A&A...472..643M}.

In this letter we study how a companion star in a binary, which formed through an exchange encounter
in a young stellar cluster, can affect solar-system-like planetary systems. 
A solar-system-like planetary system is by us defined as a planetary system in which the gas giants are 
on long-term stable orbits wider than 5 au. We compare the eccentricity distribution of the resulting systems with
the eccentricity distribution of the observed extrasolar planets at intermediate separations from their host
stars. We find that they are rather similar if most of the planetary systems originally consisted of gas 
giants with rather similar masses.

\begin{figure}
\begin{center}
\resizebox{8truecm}{!}{\includegraphics{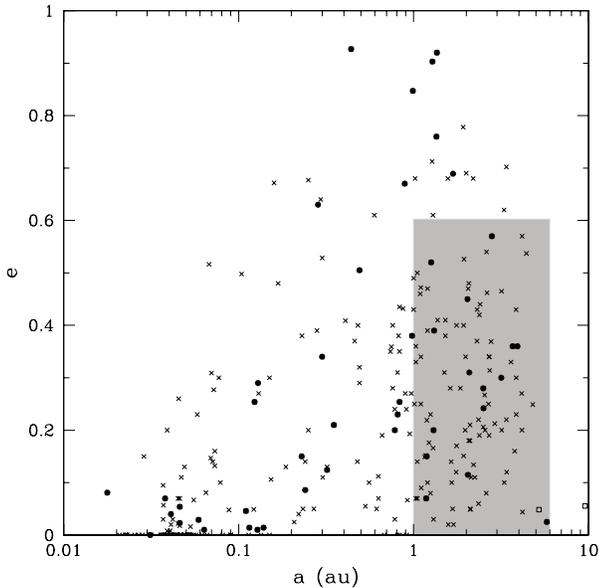}}
\caption{Eccentricity, e, of the extrasolar planets detected using the radial velocity method plotted 
against their respective semi-major axes, a (in au). The crosses identify planets orbiting single stars, 
while the filled circles identify planets in binaries. The two open squares represent Jupiter and 
Saturn. The grey region in this plot encompasses the region in phase-space, which we consider in 
this letter. }
\label{Fig:fig1}
\end{center}
\end{figure}

\section{Sample selection}
We define planets at intermediate separations as those with semi-major axes in the range 1-6 au. 
The upper limit roughly coincides with the observational limit and is only a little larger than the 
semi-major axis of Jupiter in the solar system. Hence, 6 au roughly corresponds to where we would expect 
to start finding planets in unperturbed solar-system-like planetary systems. 
From our simulations of we find that it is very unlikely for planets 
to be scattered onto orbits tighter than about 1 au. This is not an exact limit, and we do see a few 
planets on even tighter orbits in our simulations. It is however clear that the planetary systems 
produced in our simulations will essentially only contribute to the observed extrasolar planets with 
semi-major axis greater than about 1 au and hence we only compare our results to the observed 
planets with $a>1$ au. 

According to \citet{2004MNRAS.354.1165C} the detection efficiency of planets using the radial 
velocity method decreases sharply with increasing eccentricity for planets with $e>0.6$. This implies 
that the observed eccentricity distribution is not complete above an eccentricity of 0.6. However, 
according to \cite{2008arXiv0806.0032S} the decrease in detection efficiency is not as strong as that 
predicted by \citet{2004MNRAS.354.1165C} and hence the observed eccentricity distribution is a 
good reflection of the true eccentricity distribution. Nevertheless, in order to be certain that we avoid 
comparing our simulations to a biased sample we here only consider the planets in the shaded area 
of Fig. 1, hence those with semi-major axes between 1 and 6 au and with eccentricity less than 0.6. 
We have calculated the cumulative eccentricity distribution of this sample, and find it to be 
increasing approximately linearly between an eccentricity of 0 and 0.6.

\section{Planets in Binaries}
In Fig. 1 we have identified the planets found in binaries (filled circles) from those found around 
single stars (crosses) \citep{2007A&A...462..345D, 2008A&A...480L..33T}.  It is clear from the figure 
that there is no obvious difference in the eccentricity distribution for planets in binaries with respect to 
that of planets orbiting single stars for the planets in the shaded region. It is however evident that the 
four most eccentric systems are found to be in binaries. Since the number of systems is very low it is 
however too early to say if this is a real effect, or just a coincidence. 

Using a thorough statistical analysis of the observed sample of extrasolar planets, \citet{2007A&A...462..345D}
showed that there is no significant difference between the eccentricity distribution of planets in binaries and 
that of planets orbiting single stars below $e = 0.6$. Above $e = 0.6$ there is possibly a slight
excess of highly eccentric planets for the planets in (wide) binaries, although the statistical 
significance of this finding is, due to the low number of systems, not rigorous. This is a very important
result, since it leaves only two possibilities:

\begin{enumerate}

\item Planetary systems are not affected by the presence of a companion star in a binary, or

\item the perturbation from the companion star in a binary triggers the same mechanism
	as that which give rise to the observed distribution of eccentricities for planets around single stars.

\end{enumerate}

The separation of most observed planet-hosting binaries is between 100 and 1000 au and the 
companion mass is between 0.2 and 2.0 $M_{\odot}$ \citep{2007A&A...462..345D}. If most of
the binaries with $100<a<1000$ au are not primordial, but were instead formed in exchange 
encounters in young stellar clusters, the orientation of the orbital plane of the planets with 
respect to the orbital plane of the companion star is completely random  \citep[see for example]
[]{2007MNRAS.377L...1M}. In  that case about 77 per cent of the systems will have
an inclination between the orbits of the planets and the orbit of the companion star greater than $39.2\degree$. In such 
systems the so-called Kozai Mechanism \citep{1962AJ.....67..591K} operates, making the orbits 
of the planets more eccentric. This can, depending on the initial configuration of the system,
trigger strong planet-planet interactions. For example, the four giants in our solar system would,
if put inside a binary with properties similar to those observed for planet-hosting binaries, undergo
a phase of strong planet-planet interactions within a few million years, assuming that the inclination
of the companion star with respect to the planets is large enough \citep[see Fig.  2 in ][]
{2007MNRAS.378.1207M}.

Strong planet-planet interactions leave almost no trace of the initial eccentricities of the planets and 
it is thus most likely the dominating process behind the observed eccentricities among planets in 
binaries. Hence, alternative (i) is not correct, leaving us with alternative (ii): {\sl the perturbation 
from the companion star in a binary triggers the same mechanism as that which give rise to the observed 
distribution of eccentricities for planets around single stars}. This mechanism is most
likely scatterings  caused by strong planet-planet interactions. For planets which
orbit stars that are currently single, the strong planet-planet interactions might have occurred 
because the planetary system in which they formed was intrinsically unstable (see discussion in section 1). However, it 
may be that some of the observed extrasolar planets come from solar-system-like planetary systems.
In these strong planet-planet interactions occurred because the host star suffered a close encounter with 
another star or was exchanged into a binary in an exchange encounter in a young stellar cluster.

Such exchange encounters between single stars and binary systems may in fact be rather common in 
young stellar clusters, in which most stars form. We define a {\sl singleton} as a star which did not 
form in a binary, has never later spent time within a binary and has never suffered a close encounter 
with another star. Not all of the stars which are single today are singletons. N-body simulations show, 
that as a lower bound, between 5 and 10 per cent of the current single field-stars with a mass close to 
$1 \, M_{\odot}$ has previously either suffered a close encounter with another star or been part of a 
binary system \citep{2007MNRAS.378.1207M}. From a statistical analysis of radial-velocity searches 
for extrasolar planets, it has been estimated that about 7 per cent of all solar-mass stars have planets 
on orbits tighter than 5 au \citep{2005PThPS.158...24M}. Hence, external perturbations by other stars
on solar-system-like planetary systems could account for a significant fraction of the observed systems 
with separations between 1 and 6 au. It is however also
possible that the contribution is very small, depending on, for example, how common solar-system-like 
planetary systems are and how effective fly-bys are for triggering strong planet-planet interactions.

\section{Simulations}
\begin{figure}
\begin{center}
\resizebox{8truecm}{!}{\includegraphics{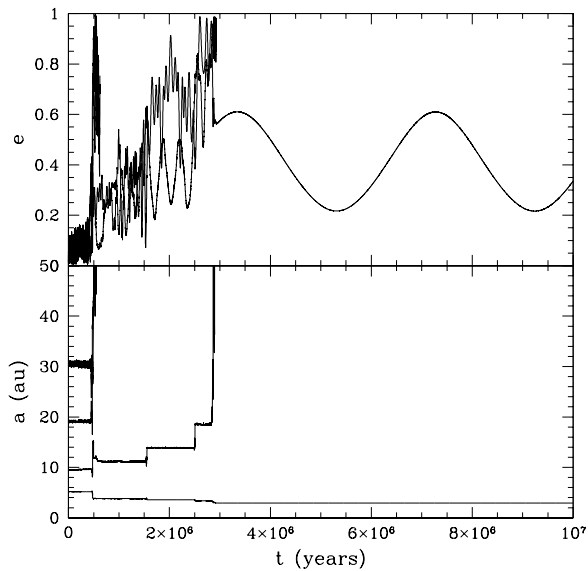}}
\caption{In the upper panel we plot the eccentricity, e, as a function of time, t (in years) for four planets
inside a binary system. The planets all had a mass equal to that of Jupiter and were initially placed on 
the same orbits as those of the gas giants in the solar system. The properties of 
the companion star were $a_{\rm c}=300$ au, $e_{\rm c}=0.3$, $i_{\rm c}=60^{\circ}$ and $m_{\rm 
c}=0.6M_{\odot}$. In the lower panel we plot the semi-major axis, a (in au) as a function of time for the 
same system. The Kozai Mechanism induces large eccentricities in the planetary orbits which leads 
to strong interactions between planets. After only 3 Myr all but one of the planets have been ejected, 
leaving that remaining on a much tighter and much more eccentric orbit.}
\label{Fig:fig2}
\end{center}
\end{figure}

We performed more than 500 simulations of several different planetary systems in binaries, using the 
publicly available MERCURY code \citep{1999MNRAS.304..793C,2002AJ....123.2884C}. All 
simulations were run for $10^{8}$ years. In all our simulations we have closely monitored the energy 
and angular momentum conservation, and if it failed using the appropriate symplectic integration 
algorithm (hybrid or wide-binary) included in MERCURY, we re-ran the simulation, using exactly the 
same initial conditions, but with a Burlish-Stoer algorithm. In the end we used the the Burlish-Stoer 
method for the majority of our simulations, since in systems where many strong close--encounters 
between planets occurred, the energy conservation using the symplectic algorithms was not good 
enough.

One would expect there to be a wide variety of planetary systems \citep{1998AJ....116.1998L} but 
here we divide them into two different groups: hierarchical and democratic. We only consider 
solar-system-like planetary systems, i.e. systems containing giant planets with separations of around 5 au 
or larger. A hierarchical system is dominated by its most-massive planet, an example of this is our 
own Solar System. A democratic system on the other hand consists of several planets of roughly 
similar, but not necessarily equal, mass. Due to current observational limits, we have not yet 
observed any multiple planet systems with planets outside 6 au, apart from our own solar system. To 
simulate a democratic system, we used the four giant planets in the solar system but set their masses 
equal to that of Jupiter. To check that this system is stable we performed several simulations of it 
around a single star and found no signs of any secular trends in the orbital elements of the planets. 
When placed in a binary, we call this system 4 Jupiters-In-Binary (4JIB). Within this system the 
perturbation of the companion star leads to large eccentricities and thus strong interactions between 
the planets, typically resulting in the ejection of all but one planet within a few Myr, leaving the 
remaining planet on a tighter and more eccentric orbit (see Fig. 2).

The effect of the companion star on the planets is to slightly perturb the outer planet, leading to strong 
planet-planet interactions in the system. Whether this happens or not depends both on the properties 
of the planetary system and on the properties of the companion star. We have kept the properties of 
the companion star constant in all the runs from which we generate eccentricity distributions below, 
with $a_{\rm c}=300 \, {\rm au}, m_{\rm c}=0.6M_{\odot} \, {\rm and} \, e_{ \rm c}=0.3$. These properties
are representative of binaries formed in exchange encounters in young stellar clusters \citep[see Fig.  5 in ][]
{2007MNRAS.378.1207M}. Furthermore,  the orientation of the orbit of the companion star with respect to 
the orbits of the planets in such binaries is random, and thus we assume that the inclination of 
binaries produced in exchange encounters is isotropically distributed.

In order to investigate how the final eccentricity distribution is affected by changing the binary 
properties we have also performed simulations of the system consisting of four Jupiter-mass planets 
in a binary, with several different values of the semi-major axis of the binary.  These show that there are 
essentially only two outcomes. Either strong planet-planet interactions are induced, resulting in an 
eccentricity distribution like that of the 4JIB with $a_{\rm c} = 300$ au or the system remains 
unperturbed, resulting in an eccentricity distribution like that of the 4J. The probability that strong 
planet-planet interactions will be induced in the four Jupiter system decreases with increasing binary 
semi-major axis. It is close to one for all inclinations up to $a_{\rm c} \simeq 800$ au and reaches 
zero for  $1000 < a_{\rm c} <1500$ au, the exact value depending on the inclination of the companion 
star. When comparing with the observed eccentricity distribution we can thus account for different 
binary properties by combining, for example, the 4J and the 4JIB distribution. Changing the mass of 
the companion star is essentially equivalent to changing its semi-major axis, since it only changes the 
strength of the perturbation. Hence we expect the same outcome to be true when varying the mass as 
when varying the semi-major axis.

In order to generate a distribution of eccentricities from our simulations we ``observe" the eccentricities 
and semi-major axes of all the planets in each simulation at several random times. To avoid the initial 
strong planet-planet interaction phase, we only considered the second half of our simulations, that is, the
last 50 Myr. We then discarded all the planets with semi-major axes smaller than 1 or larger than 6 au (the 
observational limit) and those with an eccentricity greater than 0.6. 

\section{Results}
\begin{figure}
\begin{center}
\resizebox{8truecm}{!}{\includegraphics{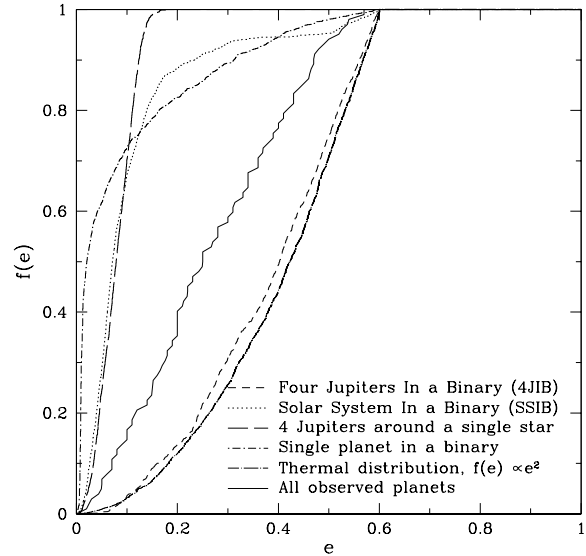}}
\caption{ The cumulative eccentricity distributions, f(e), generated from our simulations of: 4 Jupiter 
mass planets In a Binary (4JIB);  4 giants of the Solar System In a Binary (SSIB); 4 Jupiter mass 
planets around a single star (4J); a single planet in a binary. Also included is the thermal distribution 
of eccentricities and the observed extra--solar planets with separations between 1 and 6 au and 
eccentricity below 0.6. }
\label{Fig:fig3}
\end{center}
\end{figure}

We find that the democratic 4JIB system gives rise to a cumulative eccentricity distribution which goes 
roughly as $e^2$ up to 0.6. We have also performed the same set of simulations for a system where we 
instead set the mass of the four giants in the solar system equal to that of Uranus. This 
system gives rise to a very similar eccentricity distribution. It is interesting to note that the eccentricity
distribution of democratic systems is very similar to the thermal eccentricity distribution $(f(e)=e^2)$ 
found for wide stellar binaries \citep{1975MNRAS.173..729H}. We conclude that democratic systems
in general produce an excess of eccentric systems compared to the observed planets.
To compare this result with hierarchical systems, we simulated the four giants in the Solar System In 
a Binary (SSIB) and calculated the resulting eccentricity distribution. We find that there is a large 
excess of low-eccentricity systems produced compared to the observed planets.

We have also calculated the evolution of eccentricity of a single planet, considering the same stellar 
binary as used earlier, using the semi-analytic formulae derived from the so-called Kozai Mechanism 
\citep{1962AJ.....67..591K,1997AJ....113.1915I,2002Icar..158..434C}. This system can be thought of 
as an extreme example of a hierarchical system. The resulting cumulative eccentricity distribution is 
very similar to that which we found from our SSIB simulations \citep[see also][]{2005ApJ...627.1001T}. 
Planetary systems containing four Jupiters around a single star (4J) give rise to a cumulative 
eccentricity distribution having an excess of low eccentricity planets compared to observations. In Fig. 
3 we plot the cumulative distributions of all the above-mentioned systems. None of them individually 
provide a good match with the observed distribution.  The KS-probability when 
comparing the 4JIB eccentricities with those of the observed planets is about 0.001 and when 
comparing the SSIB eccentricities to those of the observed planets the KS-probability is  essentially 
zero. Hence, the eccentricities of the planets in our 4JIB and our SSIB samples are very different from
the eccentricities of the observed extrasolar planets.

\section{Discussion}
\begin{figure}
\begin{center}
\resizebox{8truecm}{!}{\includegraphics{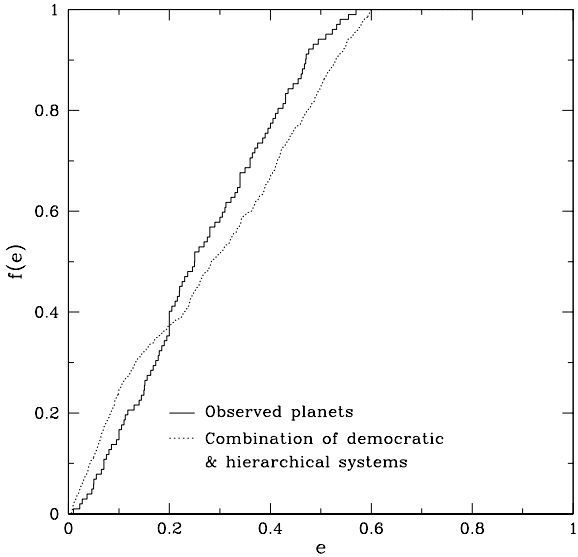}}
\caption{ The cumulative eccentricity distribution, f(e), of the observed planets with $e<0.6$ compared 
to that from a combination of 30 per cent hierarchical (solar system in binary and single planet in 
binary) and 70 per cent democratic systems  (four jupiters in binary). A KS-test of the two samples 
gives a probability of 15 per cent, which means that they are consistent with being drawn from the 
same population.}
\label{Fig:fig4}
\end{center}
\end{figure}

Since democratic systems in binaries produce an excess of eccentric systems compared to 
observations and hierarchical systems produce an excess of low eccentric systems it seems 
plausible that a combination of the two can provide a reasonable match to the observations. We plot 
an example of such a combination in Fig. 4.  In this particular case 70 per cent of the planets comes 
from the four jupiters in binary (4JIB) systems, while the remaining 30 per cent is a mix of solar system 
in binary (SSIB) and single planet in binary systems. Comparing this sample of eccentricities with the 
observed sample using a KS-test gives a probability of 0.15. This is much larger than the probability 
when using a pure sample of either democratic or hierarchical systems, and as can be seen in Fig. 4 
the match is rather good. This fit is surprisingly good considering that we have simply taken planetary 
systems derived from our own solar system. This does not suggest that all planetary systems formed
are either purely democratic or purely hierarchical. However, if the contribution to the observed 
extrasolar planet sample from solar-system-like planetary systems exchanged into (and sometimes out of) 
binaries traces the observed eccentricity distribution, the ``average" solar-system-like planetary systems 
must be significantly more democratic than the four giants in the solar system. Simulations of planetary 
formation shows that  a wide variety of planetary systems are formed, as can, for example, be seen in 
the extensive catalogue of planetary systems produced in simulations by \citet{1998AJ....116.1998L}. 
It is encouraging to note that a significant fraction of systems formed in such are, by our definition, democratic.

\section{Summary}
Most stars form in some sort of cluster or association and hence so do most planetary systems. In 
such crowded places initially single stars may be exchanged in and out of binary systems and/or 
pass close to other stars. We have performed a large set of simulations of solar-system-like planetary 
systems whose host star have been exchanged into a binary. Because the binary was formed in an 
exchange encounter the orientation of the orbits of the planets with respect to the orbit of the 
companion star is completely random. This means that the so-called Kozai mechanism operates in 
many of the systems, causing strong planet-planet interactions to occur. These in turn lead to the 
ejection of one or more planets, leaving those remaining on tighter and more eccentric orbits. We find 
that democratic planetary systems (in which the gas giants all have rather similar masses) in binaries 
produce an excess of highly eccentric systems compared to the observed extrasolar planets while 
hierarchical planetary systems (like our own solar system) in binaries produce an excess of 
low-eccentric systems. A combination of hierarchical and democratic systems in binaries does however 
provide a good match to the observed eccentricities of extrasolar planets.

\section*{Acknowledgments}
M.B.D. is a Royal Swedish Academy Research Fellow supported by a grant from the Knut and Alice 
Wallenberg Foundation. All simulations presented here were performed on computer hardware 
which was purchased with grants from the Royal Physiographic Society in Lund. We thank A. J. 
Levan, M. Mayor and G. Piotto for comments and suggestions.

\bibliography{MalmbergBib}

\begin{thebibliography}{}

\bibitem[\protect\citeauthoryear{{Adams}, {Proszkow}, {Fatuzzo} \&
  {Myers}}{{Adams} et~al.}{2006}]{2006ApJ...641..504A}
{Adams} F.~C.,  {Proszkow} E.~M.,  {Fatuzzo} M.,    {Myers} P.~C.,  2006, \apj,
  641, 504

\bibitem[\protect\citeauthoryear{{Barnes} \& {Quinn}}{{Barnes} \&
  {Quinn}}{2004}]{2004ApJ...611..494B}
{Barnes} R.,  {Quinn} T.,  2004, \apj, 611, 494

\bibitem[\protect\citeauthoryear{{Butler} et~al.}{{Butler} et~al.}{2006}]{2006ApJ...646..505B}
{Butler} R.~P. et~al.,  2006, \apj., 646, 505

\bibitem[\protect\citeauthoryear{{Carruba}, {Burns}, {Nicholson} \&
  {Gladman}}{{Carruba} et~al.}{2002}]{2002Icar..158..434C}
{Carruba} V.,  {Burns} J.~A.,  {Nicholson} P.~D.,    {Gladman} B.~J.,  2002,
  Icarus, 158, 434

\bibitem[\protect\citeauthoryear{{Chambers}}{{Chambers}}{1999}]{1999MNRAS.304.%
.793C}
{Chambers} J.~E.,  1999, \mnras, 304, 793

\bibitem[\protect\citeauthoryear{{Chambers}, {Quintana}, {Duncan} \&
  {Lissauer}}{{Chambers} et~al.}{2002}]{2002AJ....123.2884C}
{Chambers} J.~E.,  {Quintana} E.~V.,  {Duncan} M.~J.,    {Lissauer} J.~J.,
  2002, \aj, 123, 2884

\bibitem[\protect\citeauthoryear{{Chatterjee}, {Ford}, {Matsumura} \&
  {Rasio}}{{Chatterjee} et~al.}{2007}]{2007astro.ph..3166C}
{Chatterjee} S.,  {Ford} E.~B.,  {Matsumura} S.,    {Rasio} F.~A.,  2007, \apj,
  in press

\bibitem[\protect\citeauthoryear{{Cumming}}{{Cumming}}{2004}]{2004MNRAS.354.11%
65C}
{Cumming} A.,  2004, \mnras, 354, 1165

\bibitem[\protect\citeauthoryear{{Desidera} \& {Barbieri}}{{Desidera} \&
  {Barbieri}}{2007}]{2007A&A...462..345D}
{Desidera} S.,  {Barbieri} M.,  2007, \aap, 462, 345

\bibitem[\protect\citeauthoryear{{Fabrycky} \& {Tremaine}}{{Fabrycky} \&
  {Tremaine}}{2007}]{2007ApJ...669.1298F}
{Fabrycky} D.,  {Tremaine} S.,  2007, \apj, 669, 1298

\bibitem[\protect\citeauthoryear{{Ford}, {Lystad} \& {Rasio}}{{Ford}
  et~al.}{2005}]{2005Natur.434..873F}
{Ford} E.~B.,  {Lystad} V.,    {Rasio} F.~A.,  2005, \nat, 434, 873

\bibitem[\protect\citeauthoryear{{Heggie}}{{Heggie}}{1975}]{1975MNRAS.173..729%
H} {Heggie} D.~C.,  1975, \mnras, 173, 729

\bibitem[\protect\citeauthoryear{{Holman}, {Touma} \& {Tremaine}}{{Holman}
  et~al.}{1997}]{1997Natur.386..254H}
{Holman} M.,  {Touma} J.,    {Tremaine} S.,  1997, \nat, 386, 254

\bibitem[\protect\citeauthoryear{{Holman} \& {Wiegert}}{{Holman} \&
  {Wiegert}}{1999}]{1999AJ....117..621H}
{Holman} M.~J.,  {Wiegert} P.~A.,  1999, \aj, 117, 621

\bibitem[\protect\citeauthoryear{{Innanen}, {Zheng}, {Mikkola} \&
  {Valtonen}}{{Innanen} et~al.}{1997}]{1997AJ....113.1915I}
{Innanen} K.~A.,  {Zheng} J.~Q.,  {Mikkola} S.,    {Valtonen} M.~J.,  1997,
  \aj, 113, 1915

\bibitem[\protect\citeauthoryear{{Juric} \& {Tremaine}}{{Juric} \&
  {Tremaine}}{2007}]{2007astro.ph..3160J}
{Juric} M.,  {Tremaine} S.,  2007, \apj, in press

\bibitem[Kokubo 
\& Ida(1998)]{1998Icar..131..171K} Kokubo, E., \& Ida, S.\ 1998, Icarus, 131, 171 

\bibitem[\protect\citeauthoryear{{Kozai}}{{Kozai}}{1962}]{1962AJ.....67..591K}
{Kozai} Y.,  1962, \aj, 67, 591

\bibitem[\protect\citeauthoryear{{Levison}, {Lissauer} \& {Duncan}}{{Levison}
  et~al.}{1998}]{1998AJ....116.1998L}
{Levison} H.~F.,  {Lissauer} J.~J.,    {Duncan} M.~J.,  1998, \aj, 116, 1998

\bibitem[\protect\citeauthoryear{{Lin}, {Bodenheimer} \& {Richardson}}{{Lin}
  et~al.}{1996}]{1996Natur.380..606L}
{Lin} D.~N.~C.,  {Bodenheimer} P.,    {Richardson} D.~C.,  1996, \nat, 380, 606

\bibitem[\protect\citeauthoryear{{Malmberg}, {Davies} \& {Chambers}}{{Malmberg}
  et~al.}{2007a}]{2007MNRAS.377L...1M}
{Malmberg} D.,  {Davies} M.~B.,    {Chambers} J.~E.,  2007a, \mnras, 377, L1

\bibitem[\protect\citeauthoryear{{Malmberg}, {de Angeli}, {Davies}, {Church},
  {Mackey} \& {Wilkinson}}{{Malmberg} et~al.}{2007b}]{2007MNRAS.378.1207M}
{Malmberg} D.,  {de Angeli} F.,  {Davies} M.~B.,  {Church} R.~P.,  {Mackey} D.,
     {Wilkinson} M.~I.,  2007b, \mnras, 378, 1207

\bibitem[\protect\citeauthoryear{{Marcy}, {Butler}, {Fischer}, {Vogt},
  {Wright}, {Tinney} \& {Jones}}{{Marcy} et~al.}{2005}]{2005PThPS.158...24M}
{Marcy} G.,  {Butler} R.~P.,  {Fischer} D.,  {Vogt} S.,  {Wright} J.~T.,
  {Tinney} C.~G.,    {Jones} H.~R.~A.,  2005, Progress of Theoretical Physics
  Supplement, 158, 24

\bibitem[Marzari 
\& Weidenschilling(2002)]{2002Icar..156..570M} Marzari, F., \& Weidenschilling, S.~J.\ 2002, Icarus, 156, 570 

\bibitem[\protect\citeauthoryear{{Marzari}, {Weidenschilling}, {Barbieri} \&
  {Granata}}{{Marzari} et~al.}{2005}]{2005ApJ...618..502M}
{Marzari} F.,  {Weidenschilling} S.~J.,  {Barbieri} M.,    {Granata} V.,  2005,
  \apj, 618, 502

\bibitem[Marzari 
\& Barbieri(2007)]{2007A&A...472..643M} Marzari, F., \& Barbieri, M.\ 2007, \aap, 472, 643 


\bibitem[\protect\citeauthoryear{{Mayor} \& {Queloz}}{{Mayor} \&
  {Queloz}}{1995}]{1995Natur.378..355M}
{Mayor} M.,  {Queloz} D.,  1995, \nat, 378, 355

\bibitem[\protect\citeauthoryear{{Mazeh}, {Krymolowski} \& {Rosenfeld}}{{Mazeh}
  et~al.}{1997}]{1997ApJ...477L.103M}
{Mazeh} T.,  {Krymolowski} Y.,    {Rosenfeld} G.,  1997, \apj, 477, L103

\bibitem[\protect\citeauthoryear{{Moorhead} \& {Adams}}{{Moorhead} \&
  {Adams}}{2005}]{2005Icar..178..517M}
{Moorhead} A.~V,  {Adams} A.~V.,  2005, Icarus, 178, 517

\bibitem[\protect\citeauthoryear{Schneider}{Schneider}{2008}]{2008exoplanet.eu}
Schneider J., , 2008, {The Extrasolar Planets Encyclopaedia}

\bibitem[\protect\citeauthoryear{{Shen} \& {Turner}}{{Shen} \&
  {Turner}}{2008}]{2008arXiv0806.0032S}
{Shen} Y.,  {Turner} E.~L.,  2008, ApJ, {\sl in press}

\bibitem[\protect\citeauthoryear{{Takeda} \& {Rasio}}{{Takeda} \&
  {Rasio}}{2005}]{2005ApJ...627.1001T}
{Takeda} G.,  {Rasio} F.~A.,  2005, \apj, 627, 1001

\bibitem[\protect\citeauthoryear{{Tamuz} et~al.}
{{Tamuz} et~al.}{2008}]{2008A&A...480L..33T}
{Tamuz} O. et~al.,  2008, \aap, 480, L33

\bibitem[Thommes et al.(2008)]{2008ApJ...675.1538T} Thommes, E.~W., Bryden, 
G., Wu, Y., \& Rasio, F.~A.\ 2008, \apj, 675, 1538

\bibitem[\protect\citeauthoryear{{Wolszczan} \& {Frail}}{{Wolszczan} \&
  {Frail}}{1992}]{1992Natur.355..145W}
{Wolszczan} A.,  {Frail} D.~A.,  1992, \nat, 355, 145

\bibitem[\protect\citeauthoryear{{Wu}, {Murray} \& {Ramsahai}}{{Wu}
  et~al.}{2007}]{2007ApJ...670..820W}
{Wu} Y.,  {Murray} N.~W.,    {Ramsahai} J.~M.,  2007, \apj, 670, 820

\bibitem[\protect\citeauthoryear{{Zakamska} \& {Tremaine}}{{Zakamska} \&
  {Tremaine}}{2004}]{2004AJ....128..869Z}
{Zakamska} N.~L.,  {Tremaine} S.,  2004, \aj, 128, 869

\end{thebibliography}
\bibliographystyle{mn2e} 
\label{lastpage}
\end{document}